\documentclass{vgtc}                          

\ifpdf
  \pdfoutput=1\relax                   
  \usepackage{graphicx}                
  \DeclareGraphicsExtensions{.pdf,.png,.jpg,.jpeg} 
\else
  \usepackage{graphicx}                
  \DeclareGraphicsExtensions{.eps}     
\fi%

\graphicspath{{figures/}{pictures/}{images/}{./}} 

\usepackage{microtype}                 
\PassOptionsToPackage{warn}{textcomp}  
\usepackage{textcomp}                  
\usepackage{mathptmx}                  
\usepackage{times}                     
\usepackage{cite}                      
\usepackage{tabu}                      
\usepackage{booktabs}                  
\usepackage{multirow}
\usepackage{amsfonts} 
\usepackage{amssymb}
\usepackage{amsmath}
\usepackage{color}

\newcommand{\zlz}[1]{\textcolor{black}{#1}}

\onlineid{0}

\vgtccategory{Research}

\vgtcinsertpkg

\title{In-Place Gestures Classification via\\ Long-term Memory Augmented Network}

\author{
Lizhi Zhao \thanks{Joint first author.} \\ \scriptsize Northwest A\&F University
\and Xuequan Lu \footnotemark[1] \\\scriptsize Deakin University
\and Qianyue Bao \\\scriptsize Xidian University %
\and Meili Wang  \thanks{Corresponding author: wml@nwsuaf.edu.cn \newline \hspace*{1.8em}Key Laboratory of Agricultural Internet of Things, Ministry of Agriculture, Yangling 712100, China \newline \hspace*{1.5em} Shaanxi Key Laboratory of Agricultural Information Perception and Intelligent Service, Yangling 712100, China 
\newline \hspace*{1.5em} This work was partially funded by 2021 Science and Technology Innovation Program of Shaanxi Academy of Forestry Science (SXLK2021-0214). 
} \\ \scriptsize Northwest A\&F University 
}

\abstract{
In-place gesture-based virtual locomotion techniques enable users to control their viewpoint and intuitively move in the 3D virtual environment.
A key research problem is to accurately and quickly recognize in-place gestures, since they can trigger specific movements of virtual viewpoints and enhance user experience.    
However, to achieve real-time experience, only short-term sensor sequence data (up to about 300ms, 6 to 10 frames) can be taken as input, 
which actually affects the classification performance due to limited spatio-temporal information.  
In this paper, we propose a novel long-term memory augmented network for in-place gestures classification. It takes as input both short-term gesture sequence samples and their corresponding long-term sequence samples that provide extra relevant spatio-temporal information in the training phase. We store long-term sequence features with an external memory queue. In addition, we design a memory augmented loss to help cluster features of the same class and push apart features from different classes, thus enabling our memory queue to memorize more relevant long-term sequence features.
In the inference phase, we input only short-term sequence samples to recall the stored features accordingly, and fuse them together to predict the gesture class.
We create a large-scale in-place gestures dataset from 25 participants with 11 gestures. 
Our method achieves a promising accuracy of 95.1\% with a latency of 192ms, and an accuracy of 97.3\% with a latency of 312ms, and is demonstrated to be superior to recent in-place gesture classification techniques. 
User study also validates our approach. 
\textit{Our source code and dataset will be made available to the community. }
} 

\CCScatlist{
  \CCScatTwelve{Human-centered computing}{Human computer interaction (HCI)}{Interaction techniques}{Gestural input};
  \CCScatTwelve{Human-centered computing}{—Human computer interaction (HCI)}{Interaction paradigms}{Virtual reality}
}

\begin{document}



\maketitle

\section{Introduction}
Locomotion in virtual environment refers to controlling the user's viewpoint movement in the 3D environment, which is a basic and common interaction technique for Virtual Reality (VR) applications \cite{bowman1998vrtravel}. 
In recent years, with the development of motion-capable VR devices such as head-mounted displays (HMDs), controllers, and trackers, gesture-based virtual locomotion techniques have received considerable attention, which allows users to navigate the virtual environment with real walking or walking-like gestures and provides kinesthetic feedback and improve immersion and naturalness \cite{nilsson2016walking}. 

In small or cluttered rooms, users can only move safely in a small restricted area or stay in place, hence many studies have focused on \textit{in-place} gestures for virtual locomotion such as walking in place (WIP) \cite{zhao2021classifying,shi2019accurate,wendt2010gud,hanson2019improving}, jumping \cite{redirectedjumping,wolf2020jumpvr}, body leaning \cite{buttussi2019locomotion}, arm swinging \cite{pai2017armswing}, etc.  
These in-place gesture-based locomotion methods concentrate on recognizing specific pre-defined gesture patterns from motion sequence data collected by motion capture sensors (i.e., HMDs, trackers, etc.) with low latency and high accuracy, for controlling the direction and speed of virtual locomotion. 
Walking in place (WIP) is a partial gait locomotion method that enables users to direct their forward movement in virtual environment by stepping in place \cite{templeman1999wip}, which is an inexpensive way to provide proprioceptive feedback similar to real walking \cite{al2018locomotionsurvey,nilsson2018natural}.

Traditional WIP methods typically design gait patterns such as threshold parameters manually to recognize WIP gestures \cite{wendt2010gud,feasel2008llcm}.
These methods highly rely on prior knowledge, and can only recognize a few types of gestures such as standing and walking while ignoring other in-place gestures (i.e., jogging or jumping). They also have poor generalization to different users. 
In-place gesture-based locomotion methods using deep/machine learning have been proposed recently to solve the above problem. 
For example, Hanson \textit{et al.} \cite{hanson2019improving} proposed to treat WIP as a classification task and trained a simple convolutional neural network to recognize walking and standing gestures using sensor sequence data as input. 
Shi \textit{et al.} \cite{shi2019accurate} proposed DCTC, an improved LSTM network, which takes as input the pressure sensor signals for recognizing 7 gait patterns. 
Zhao \textit{et al.} \cite{zhao2021classifying} treated several consecutive frames of 3D sensor data as a ``point cloud'' and extracted gesture features with a point cloud learning model. In addition, they employed an unsupervised domain adaptation method to bridge the domain gap between users. 
However, as described in \cite{zhao2021classifying,shi2019accurate}, these deep learning-based methods require sequence duration to collect sensor data as input samples during the real-time inference phase, which leads to latency. 
With a larger sequence duration, the sensor sequence data can provide richer geometric trajectory shape \cite{guo20143d,guo2013rotational} and spatio-temporal information of the user's gestures, and the network tends to produce higher accuracy with higher latency.

We propose a novel Long-term Memory Augmented Network (LMAN) for real-time in-place gestures classification, motivated by the observation that human action usually lasts for a period of time, and the short-term sequence of the entire action is semantically relevant to the long-term sequence (containing more rich spatio-temporal information) encompassing it. 
In the training phase, we input both short-term sequence samples and their corresponding long-term sequence samples into the network.
Our LMAN involves an external memory queue to store long-term sequence features, which can be recalled by short-term sequence features through similarity matching to provide extra relevant gesture contextual information.
During the inference phase, we input only short-term sequence samples, which are fused with the recalled relevant long-term sequence features and fed into the decoder to predict gestures.
In addition, we propose the memory augmented loss (MAL) to encourage LMAN to memorize more relevant and robust features by clustering short-term sequence features with the same class of long-term sequence features in the memory queue and pushing apart features from different classes simultaneously. 
The contributions of this paper are summarized as follows:
\begin{itemize}
\item We propose a novel Long-term Memory Augmented Network with an external memory queue to store long-term gesture sequence features, which can be recalled by short-term sequence features to provide extra rich spatio-temporal information for real-time in-place gestures classification. To our knowledge, it is \textit{the first work} for in-place gestures classification with Memory Augmented Networks.
\item We propose the memory augmented loss to drive LMAN to memorize more robust and relevant long-term sequence features, which improves the classification accuracy of the model.
\item We build a large-scale dataset including 11 in-place gestures from 25 participants, containing a total of 1,571,069 labeled frames, which is currently the largest open dataset for in-place gestures.
\textit{We will open our dataset and source code to the community.}
\item We conduct comprehensive experiments on our dataset. Our method achieves a promising accuracy of 95.1\% with a latency of 192ms, and an accuracy of 97.3\% with a latency of 312ms, and is shown to be superior to recent in-place gesture classification techniques.
User study also confirms the effectiveness and responsiveness of our approach.
\end{itemize}


\section{Related Work}
\subsection{In-place Gesture-based Virtual Locomotion}
Virtual locomotion is to control the user's viewpoint for moving in a 3D virtual environment while keeping the user in a relatively small physical space \cite{templeman1999wip}. 
Joystick-based virtual locomotion methods are widely used in video games, allowing users to push or press the controller joysticks to move their viewpoint \cite{2019locomotionJTL,jaeger2001comparison}. 
However, these methods tend to cause motion sickness in virtual reality environments due to the inconsistency between the user's physical and visual perceptions. 
Teleportation techniques allow users to point to the destination and then move instantly with the controllers, thus avoiding motion sickness, but they provide a relatively low sense of presence and immersion \cite{2019locomotionJTL,bozgeyikli2016point,christou2017steering,langbehn2018evaluation}. 
In-place gesture-based virtual locomotion methods such as arm swing  \cite{pai2017armswing,wilson2016vr}, WIP \cite{ke2021larger,hanson2019improving,shi2019accurate,zhao2021classifying,tregillus2016vrstep} and body leaning \cite{2019locomotionJTL,nguyen2019naviboard} increase naturalness and immersion by mimicking full or partial movements of real walking \cite{al2018locomotionsurvey}. 
In this work, we focus on virtual locomotion methods with in-place leg/foot gestures. These methods free both hands for interacting with virtual objects and can provide better spatial awareness than arm-swinging methods \cite{wilson2016vr}.

In-place gestures for virtual locomotion are typically detected by wearable sensors such as head-mounted displays, and inertial measurement units (IMUs) in smartphones and trackers, providing spatio-temporal information (i.e., 3D position coordinates, rotation angle, velocity, etc) of body and limbs \cite{hanson2019improving}. 
VR-STEP \cite{tregillus2016vrstep} used IMUs to capture the acceleration signals of gestures, which were input to the dynamic threshold-based real-time step detection algorithm proposed by Zhao \cite{zhao2010full}. When a step was detected, VR-STEP translated it into virtual locomotion in the direction of the user's gaze. 
Based on the prior knowledge of real walking biomechanics, Wendt \textit{et al.} \cite{wendt2010gud} proposed a Gait Understanding-Driven WIP model with manual-tuning parameters to measure step frequency with only a fraction of a completed step, thus yielding low start-stop latency. 
Jung \textit{et al.} \cite{redirectedjumping} predefined the jumping cycle into five phases called Idle, Ready, Up, Down, and Landing, with each phase corresponding to an individual virtual locomotion rule. 
Users transitioned between 5 jumping phases based on tracked head, wrist, and foot positions. 
These methods require handcrafted pattern features and rely heavily on empirical knowledge of gestures. 

\subsection{Deep Learning for Virtual Locomotion}
Accurate recognition of specific gestures is crucial for gesture-based virtual locomotion.  Recently deep learning techniques have been introduced to solve this problem. 
Hanson \textit{et al.} \cite{hanson2019improving} proposed to consider WIP as a classification task for standing and walking and trained a simple convolutional neural network model to classify these two gestures. 
They sampled the acceleration signals of the user's head with HMD, which were grouped into time-series samples and fed into the model. 
Shi \textit{et al.} \cite{shi2019accurate} proposed a Dual-Check Till Consensus (DCTC) model for  the classification of seven feet gestures, which can dynamically adjust the sequence duration of the input time series sensor data with respect to classification confidence. 
Ke \textit{et al.} \cite{ke2021larger} trained the Support-Vector-Machine classification model for speed control of virtual locomotion based on tracker data of three in-place leg gestures. 
Paik \textit{et al.} \cite{paik2021backward} focused on forward and backward gestures and considered three sensor data sources (i.e., head, waist, and foot movements) and collected the corresponding sensor position data to build a dataset. To eliminate the impact of unconscious shifting when users walk in place, they re-adjusted users' position to the initial center point. 
They trained a BiLSTM model to recognize these two gestures. 
\zlz{These works typically treat sensor sequence data as two-dimensional temporal sequences, and employ simple machine learning models to classify gestures with very few classes.}

\zlz{
PointNet \cite{pointnet} is a pioneer in directly consuming point clouds for feature learning and obtains the permutation invariance of points with a symmetric function. 
Qi \textit{et al.} further proposed PointNet++ \cite{pointnet++} for capturing the local structures of each point's neighborhood by a hierarchical network consisting of PointNet modules. 
}
Zhao \textit{et al.} \cite{zhao2021classifying} treated several consecutive frames of 3D sensor data as a ``point cloud'' and extracted gesture features with a point cloud learning model.  
They also suggested that domain gap does exist between users due to inter-person variations (i.e., differences in height, weight, gender, exercise habits, etc.), leading to trained models effective for some users but less effective for others. 
They developed an end-to-end joint framework consisting of a supervised point cloud learning module to extract point cloud features, and an unsupervised domain adaptation module to bridge the domain gap between users.
\zlz{
Palipana \textit{et al.} \cite{Pantomime} introduced a  mid-air gesture recognition model which takes sparse 3D point clouds from radar sensor signals as input and combines PointNet++\cite{pointnet++} with LSTM modules to extract frame-wise spatio-temporal features.
However, the point cloud inherently lacks topological information \cite{dgcnn}.
Therefore, treating consecutive frames of motion sequence data as a point cloud hides the temporal features such as motion velocity and acceleration.
}

 
Yan \textit{et al.} \cite{stgcn} proposed a novel spatio-temporal graph convolutional network (ST-GCN), which constructed a spatio-temporal graph for skeletons, with skeleton joints as graph nodes, and edges constructed from skeleton structure and consecutive frames, respectively. In addition, they designed graph convolution kernels  to learn the higher-level features of the spatio-temporal graph. Song \textit{et al.} \cite{song2021efficientgcn} embedded the separable convolutional layers into the Multiple Input Branches (MIB) network and designed a scaling strategy to obtain the EfficientGCN models with high accuracy and small amounts of parameters for action recognition. In this work, we treat gesture-based virtual locomotion as a \textit{real-time} skeleton-based action recognition problem, with wearable sensors providing 3D position coordinates of skeleton joints.

\subsection{Memory Augmented Networks}
Memory augmented networks have been proposed to solve various computer vision tasks, such as image generation \cite{wu2018memory,jeong2021memory}, person re-identification \cite{eom2021video}, few-shot learning \cite{cai2018memory}, video prediction \cite{lmcmemory}, video object detection \cite{chen2020memory,kim2021robust}, trajectory prediction \cite{marchetti2020mantra}, and so on.
Memory augmented networks use a controller module with external element-wise addressable memory slots to store additional information, which can be selectively accessed by relevant items  \cite{marchetti2020mantra}. 
Lee \textit{et al.} \cite{lmcmemory} preserved long-term motion contexts of training data using a memory module with external independent parameters for predicting future frames with short-term motion sequence input. These stored long-term motion contexts can be recalled from the input short-term sequence. 
Kim \textit{et al.} \cite{kim2021robust} proposed a Large-scale Pedestrian Recalling (LPR) Memory to memorize the visual features of large-scale pedestrians, which can be then recalled by insufficient small-scale pedestrian appearances through relevant information addressing.

\section{Method}

\subsection{Data Collection}
\label{section:datacollection}
We mainly follow Zhao \textit{et al.}'s \cite{zhao2021classifying}  data collection setting.
We used the HTC VIVE Pro HMD and two VIVE trackers attached to the front of the participants' left and right thighs to collect 6 DoF head and leg 3D position coordinates.
We acquire the 11 in-place gestures dataset from 25 participants, including standing, walking in place, jogging in place, jumping, squatting, stepping forward, stepping backward, stepping left, stepping right, sitting, and sitting-marching \cite{ke2021larger}, with each gesture lasting two minutes. Our dataset contains a total of 1,571,069 frames.
The 25 participants are from a local university, with an average age of 23.0 years old and a standard deviation of 1.6. There are 7 females and 18 males in them. 
We investigate participants' familiarity with VR using the questionnaire from \cite{shi2019accurate}. The average familiarity score is 2.6 and the standard deviation is 1.2. 

We manually annotate each frame of the dataset with its corresponding gesture. We use the sliding window method to split the entire sequence data into \(N\) skeleton-based short-term gesture sequence samples \(\mathcal{X}=\left\{x \in \mathbb{R}^{C \times T \times V}\right\}\), where \(C=3\) denotes 3 dimensions, and \(T\) denotes the number of short-term sequence length in frames.  \(V\) denotes skeleton joints (i.e., the number of tracking sensors, in this paper \(V = 3\), indicating the HMD and two trackers). 	
For each window containing more than one gesture class, we re-slide it until the window contains only one gesture class to avoid label ambiguity. 
Since each split short-term sequence sample contains only one gesture class, we use that gesture as the label of the sample. We denote labels of the sequence samples as \(\mathcal{Y}\).

\subsection{Gesture Classification}
Long-term gesture sequence samples (i.e., samples with a large number of frames) provide more spatio-temporal information and thus can facilitate the classification network to achieve higher accuracy \cite{zhao2021classifying,shi2019accurate}. 
However, during the real-time gesture inference phase, generating long-term sequence samples from sensors requires a large sequence duration, which leads to system latency and reduces user experience. 
Therefore, in the inference phase, we can only use short-term sequences (about 6 to 10 frames) as input, which in turn limits the performance of the network. 

To solve the above problem, we propose a novel Long-term Memory Augmented Network which inputs both short-term sequence samples and their corresponding long-term sequence samples in the training phase, and stores long-term sequence features with an external memory queue. These stored features can be recalled to provide extra spatio-temporal information to complete short-term sequence samples. 
The introduced LMAN is driven by the motivation that human action usually lasts for a period of time, and the short-term sequence of the entire action is highly semantically relevant to the long-term sequence (which contains more context information) encompassing it.

In addition, we design a memory augmented loss to help pull short-term sequence features to be close to its same class long-term sequence features in embedding space, thus enabling our LMAN to memorize more relevant long-term sequence features. We introduce MoCo \cite{he2020momentum} optimization strategy to facilitate loss convergence.



\subsubsection{Long-term Memory Augmented Network}

\begin{figure*}[h] 
\centering 
\includegraphics[scale=0.6]{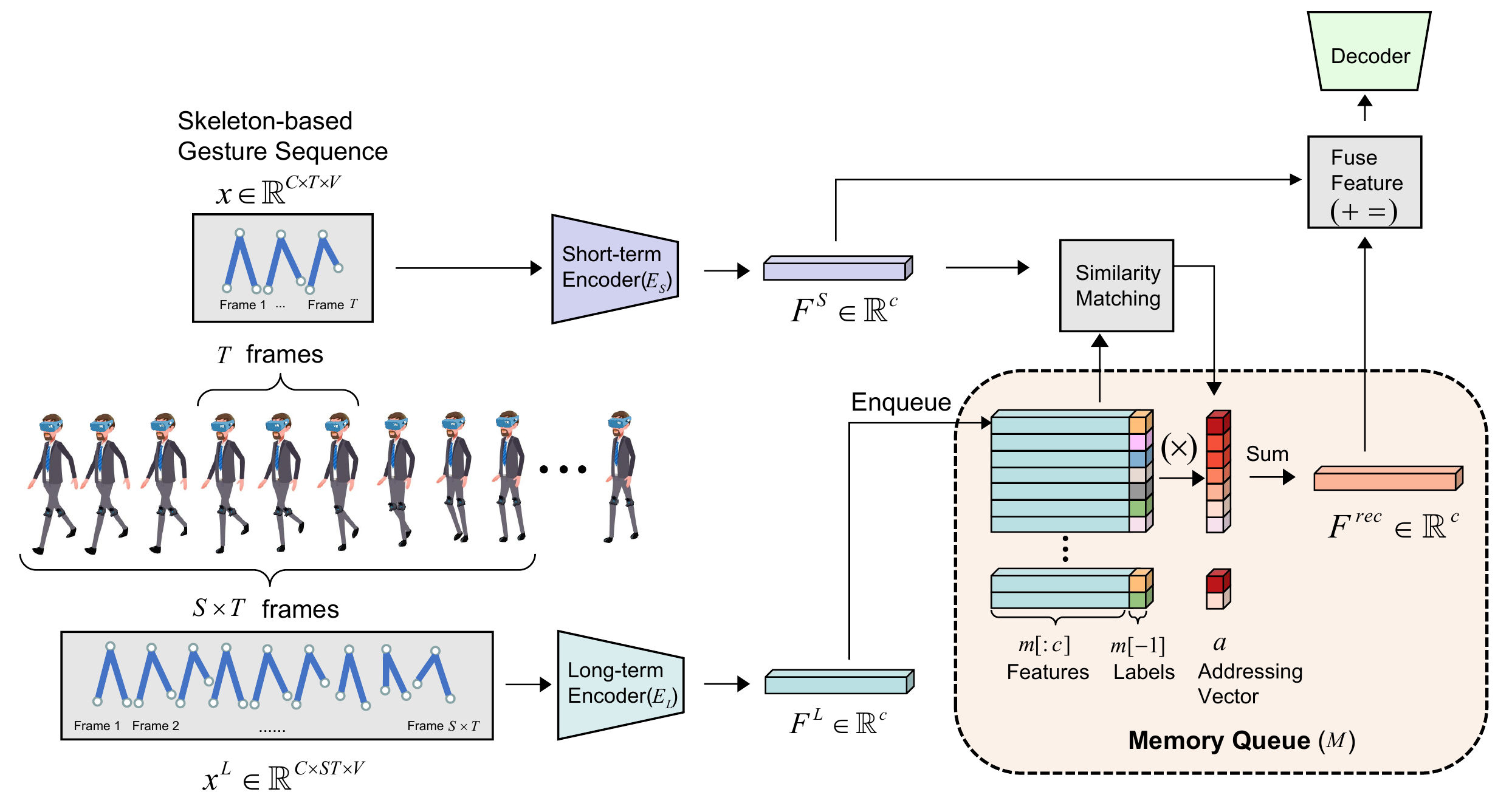} 
\caption{Our LMAN architecture. 
We capture skeleton-based gesture sequence data using HMD and two VIVE trackers. 
In training, we input both short-term sequence samples with \(T\) frames and their corresponding long-term sequence samples with \(S\times T\) frames into \(E_S\) and \(E_L\), respectively. 
We store long-term sequence features \(F^L\) in \(M\), which can be recalled by short-term sequence features \(F^S\) through similarity matching to provide extra relevant spatio-temporal feature \(F^{rec}\) for \(F^S\). 
\((\times)\) denotes element-wise multiplication, and \((+=)\) denotes element-wise add of two vectors with the same dimension. Sum denotes summing matrix elements over the row.  
} 
\label{fig:netarch} 
\end{figure*}
\vspace{-0.18cm}

Figure \ref{fig:netarch} shows the overall architecture of the proposed LMAN.
Given a short-term gesture sequence sample \(x_i\) representing the $i$-th sample of \(\mathcal{X}\) and its label \(y_i\), we denote \(x_i^L=\mathcal{X}_{i-\hat{S}:i+S-\hat{S}}=\left\{x_{t}\right\}_{t=i-\hat{S}}^{i+S-\hat{S}} \in \mathbb{R}^{C \times ST \times V}\) as the concatenation of contiguous short-term  sequence samples centered at \(x_i\), where \(\hat{S} =\lfloor S / 2 \rfloor  \), and \(S\) is the window scaling factor controlling the length of long-term sequence.
With input \(x_i\) and its corresponding long-term sequence sample \(x_i^L\), our goal is to optimize the classification function \(\mathcal{F}\) to estimate \(p(\hat{y_i}|x_i,x_i^L)\), where \(\hat{y_i}\) is the predicted label of \(x_i\).

We use the short-term encoder \(E_S\) and long-term encoder \(E_L\) to project the input long-term sequence sample \(x_i^L\) and short-term sequence sample \(x_i\) into two separate latent representations (or features) \(F_i^S=E_S(x_i) \in \mathbb{R}^c \)  and \(F_i^L=E_L(x_i^L) \in \mathbb{R}^c\), where \(c=128\) denotes the feature dimension.
\(E_S\) and \(E_L\) share the same network structure but with independent optimizable parameters. 

An external memory queue \(M\) is employed to provide extra long-term spatio-temporal information for the input short-term sequence samples. 
\(M\) is represented as matrix form \(M=\left\{m_{j}\right\}_{j=1}^{K} \in \mathbb{R}^{K \times (c+1)}\), with \(K\) memory slots and \(c+1\) channels for storing latent representation (i.e., \( m[:c]\)) and its corresponding label (i.e.,  \(m[-1]\)).
We first store the detached long-term latent representation \(F_i^L\) and \(y_i\) into the memory queue under the first-in-first-out rule. The back-propagation gradients are truncated before \(M\) as described in Section \ref{memorylearningprocedure}. 

Then the short-term sequence feature \(F_i^S\) is used as a query to match with \(M\) to recall relevant long-term sequence features. 
Following previous works \cite{lmcmemory,marchetti2020mantra,eom2021video}, we compute normalized cosine similarity between query \(F_i^S\) and all memory slots in \(M\) to produce the memory addressing vector \(a_{i}=\left\{a_{i_{-}j}\right\}_{j=1}^{K} \in \mathbb{R}^{K}\) for query \(F_i^S\), where \(a_{i_{-}j}\) can be formulated as:
\begin{equation}
a_{i_{-}j}=\frac{\exp \left( \left( F_i^S \right)^\mathrm{T} m_{i}[:c]\right)}{\sum_{j=1}^{k} \exp \left(\left( F_i^S \right)^\mathrm{T} m_{j}[:c]\right)}
\label{addressing}
\end{equation}
The memory addressing vector \(a_i\) can be considered as the attention weight \cite{lmcmemory} for each memory slot \(m_i\) in \(M\)  for producing the recalled feature \(F_i^{rec}\):
\begin{equation}
F_{i}^{rec}=\sum_{j=1}^{s} a_{i_{-}j} m_{j}[:c]
\label{weightaverage}
\end{equation}
Finally we fuse the short-term sequence feature \(F_i^S\) and \(F_i^{rec}\) and feed it into the decoder which classifies the fused feature into \(N_c\) classes (11 kinds of gestures) to obtain the \(N_c\)-dimensional probabilistic output \(p(\hat{y_i}|x_i,x_i^L)\).


\subsubsection{Memory Augmented Loss}
\label{section:MAL}
Humans tend to repeat their motion, not only for periodical actions like walking but also for other more complex actions \cite{hisrep}. Thus the split sub-sequences of a complete action are highly relevant to each other in terms of temporal-spatial information. 
For example, when a human walks, the movements of the left leg and the right leg are symmetrical within two consecutive steps, and the movements of the same leg are repetitive among steps. 
Therefore, for the input short-term sequence sample, we expect its latent representation to be clustered with the same class of long-term sequence latent representations in \(M\) while pushing apart representations from different classes simultaneously. 
Inspired by the supervised contrastive loss (SCL) \cite{supcon}, we introduce a memory augmented loss to achieve this:
\begin{equation}
\mathcal{L}_{\text {aug }}=\sum_{i \in I} \frac{-1}{|P(i)|} \sum_{p \in P(i)} \log \frac{\exp \left(F_i^S \cdot m_p[:c] / \tau\right)}{\sum_{a \in A(i)} \exp \left(F_i^S \cdot m_a[:c] / \tau\right)}
\end{equation}
We denote \(i \in I \equiv\{1 \ldots N\}\) as the index of samples in  \(\mathcal{X}\), and  \(d \in D \equiv\{1 \ldots K\}\) as the index of memory slots in \(M\). 
\(P(i) \equiv\left\{p \in D: m_p[-1]=y_{i}\right\}\) is the set of indices of memory slots in \(M\) with the same class of \(x_i\), and \(|P(i)|\) is its cardinality.
In contrast to \(P(i)\), we define \(A(i) \equiv\left\{a \in D: m_a[-1] \neq y_{i}\right\}\) as the set of indices of memory slots 
with different classes of \(x_i\).
\(\tau \in \mathbb{R}^{+}\) is a scalar temperature parameter.

Finally, we combine the cross-entropy loss \(\mathcal{L}_\text {c}\) and the memory augmented loss \(\mathcal{L}_\text {aug}\)  as our loss function \(\mathcal{L}\):
\begin{equation}
\mathcal{L} = \mathcal{L}_\text {c} + \mathcal{L}_\text {aug}
\end{equation}


\subsubsection{Memory Learning Procedure}
\label{memorylearningprocedure}
In the training phase, we input both the short-term sequence sample \(x_i\) and its corresponding long-term sequence sample \(x_i^L\) to LMAN and store the detached long-term sequence feature \(F_i^L\) in \(M\). 
Since a large number of long-term sequence features are stored in \(M\), updating parameters for \(M\) and \(E_L\) by back-propagation will cause huge computation.  
Inspired by MoCo \cite{he2020momentum}, we ignore the gradients of \(M\) and \(E_L\), and only \(E_S\) is updated by back-propagation, while \(E_L\) is updated by momentum as: 
\begin{equation}
\theta_L \leftarrow v \theta_{L}+(1-v) \theta_{S}
\end{equation}
where \(\theta_L\) denotes the parameters of \(E_L\), and  \(\theta_S\) denotes the parameters of \(E_S\), and \(v \in[0,1)\) is the momentum coefficient.
\(M\) is updated in each epoch simply by enqueuing new long-term sequence feature and dequeuing the oldest feature. 

In the test phase, we input only a short-term sequence sample \(x_i\) to recall relevant long-term sequence features \(F_i^{rec}\)  from \(M\), which is fused with \(F^S_i\) and then fed into the decoder to predict the class label of \(x_i\).

\section{Experiments}

\subsection{Implementation Details}
\zlz{Following previous works \cite{nturgb+d,shi2019accurate, ke2021larger}, we also use 
the cross-subject evaluation by dividing all subjects into training and testing sets.}
We select the gesture data of 7 participants (2 females, 5 males) from all 25 participants as the test set and the gesture data of the remaining 18 participants (5 females, 13 males) as the training set. 
\zlz{
We consider gender, height, age, weight, and familiarity with VR to select participants for testing. 
We ensure that these features are as widely distributed as possible in the test set to verify the model's generalizability. 
}

We implement our method on PyTorch. We employ Efficient-B0 \cite{song2021efficientgcn} as our encoder. 
Our model is trained by a stochastic gradient descent optimizer \cite{bottou2010large} with a weight decay of 0.0001 and a learning rate of 0.005. 
Following MoCo \cite{he2020momentum}, the memory slot size \(K\) is set to 65536, and momentum \(v\) is set to 0.99. 
The temperature parameter \(\tau\) is set to 0.07.
We trained our model with a batch size of 64 for 50 epochs on an NVIDIA RTX 3060Ti GPU, taking approximately 2.5 hours. 
The inference time of our model for each sample is approximately 12ms.

\subsection{Model Evaluation}
\subsubsection{Comparisons of Sequence Length}
As described in Section \ref{section:datacollection}, we used the sliding window method to split the entire dataset into skeleton-based short-term gesture sequence samples with corresponding long-term sequence samples.
In this section, we compare our model's gesture classification accuracy for different combinations of short-term and long-term sequence length in frames. 
We use \(l_s\)-\(l_l\) to denote the settings of the short-term sequence length and the corresponding long-term sequence length.

\begin{figure}[htbp]
\centering
\includegraphics[width=1.0\linewidth]{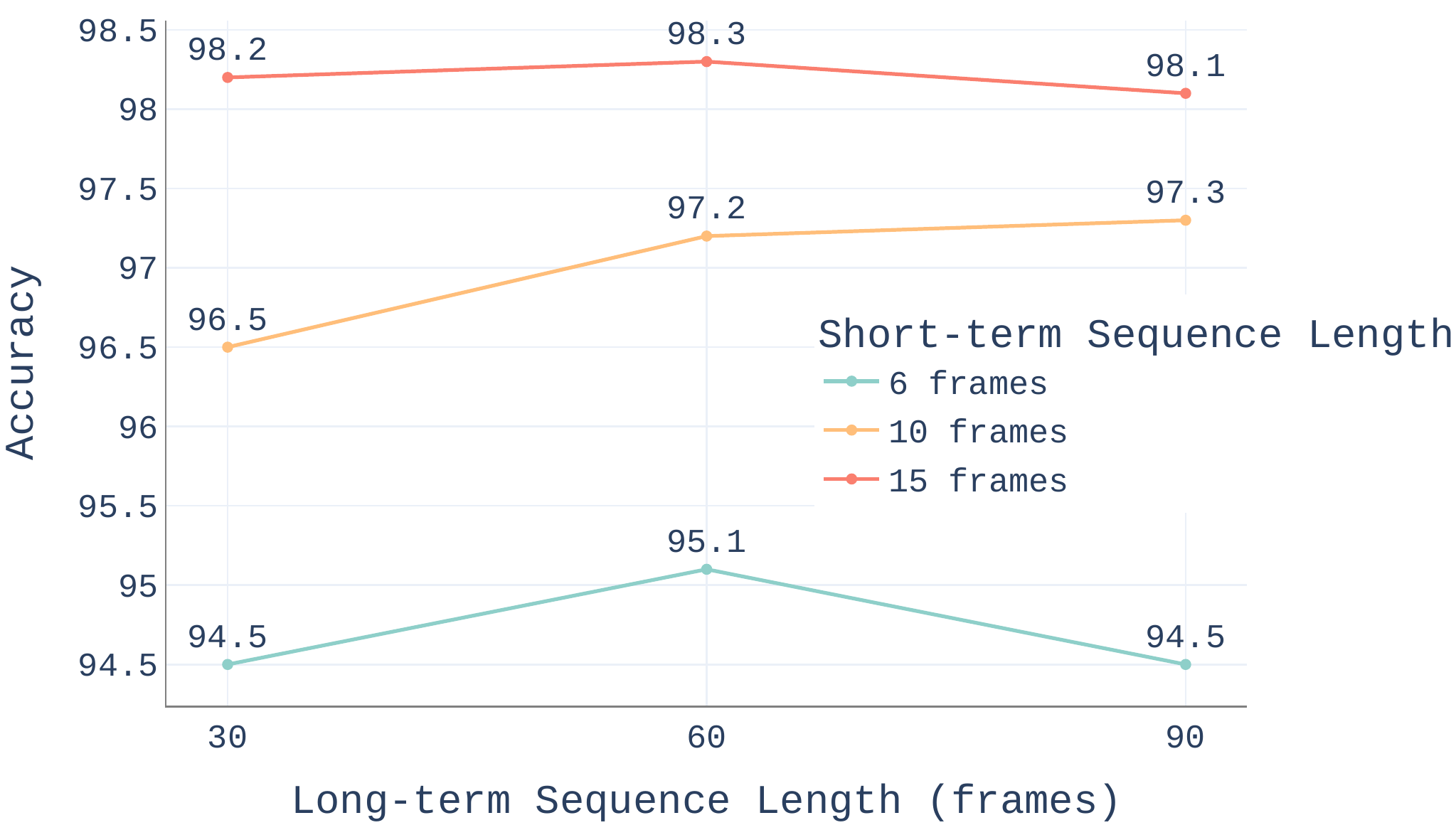}
\caption{
Comparison of model accuracy with different long-term sequence length for short-term sequence length (6, 10 and 15 frames). 
}
\label{fig:longlen}
\end{figure}

From Figure \ref{fig:longlen}, our model achieves accuracies of 95.1\%, 97.3\% and 98.3\% for short-term sequences length of 6, 10, and 15 frames, respectively. 
Note that our sensor sampling frequency is 30Hz (i.e., 30ms per frame), and the model inference time is about 12ms, which indicates that our model can achieve 95.1\% accuracy with a latency of only 192ms (6 \(\times\) 30ms + 12ms) using the 6-60 setting, which is sufficient for real-time gesture inference  \cite{zhao2021classifying}. 
In addition, the 10-90 setting yields a higher accuracy of 97.3\% with 312ms (10 \(\times\) 30ms + 12ms) latency. 

Figure \ref{fig:longlen} also demonstrates that although the difference in short-term sequence length is only 4-5 frames, larger short-term sequence length can yield higher accuracy under three long-term sequence length settings, suggesting that larger short-term sequence samples contain richer spatio-temporal gesture features that could improve gesture classification. 
With the long-term sequence length setting to 60 frames, all three short-term sequence length settings achieved relatively high accuracies, suggesting that human gestures lasting about 1.8s (60 \(\times\) 30ms) possess enough discriminative features. But gestures lasting longer (i.e., 90 frames) may contain irrelevant movements and may confuse the model.

\subsubsection{Comparisons of Memory Augmented Loss and Supervised Contrastive Loss}
As described in Section \ref{section:MAL}, we propose a memory augmented loss. 
The major difference between our MAL and SCL is the definition of positive pairs and negative pairs which are essential in contrastive learning. 
SCL randomly selects \(N_b\) samples with two random data augmentations, and thus generates \(2N_b\) augmented samples as multiviewed batch \(B_m\).
For a certain augmented sample \(\hat{x_i}\) and its label \(\hat{y_i}\), SCL treats samples with the same class of \(\hat{y_i}\) in \(B_m\) as positive pairs, 
and the \(2N_b-1\) samples except \(\hat{x_i}\) in \(B_m\) as negative pairs.
In contrast, our MAL treats memory slots in \(M\) with the same class of input short-term sequence samples as positive pairs (denoted as \(P(i)\)), but treats slots with the different classes as negative pairs (denoted as \(A(i)\)), which takes class information into account for defining negative pairs. 


\zlz{
We compare MAL with SCL,
and the results in Table \ref{tab:negative}
show that our MAL achieves about 0.5\% improvement under all three dataset settings.
To verify the significance of improvements, we conduct two-tailed t-tests on three groups of experiments trained with MAL under 6-60, 10-60 and 10-100 dataset settings.
The t-test results demonstrate that the classification accuracy of MAL significantly exceeds the accuracy of SCL 
under 6-60 dataset setting (t = 3.236, p = 0.048) and under 10-60 dataset setting (t = 3.382, p = 0.043), and under 10-100 dataset setting (t = 3.363, p = 0.044), since their p-values are all less than 0.05.
}


\begin{table}
\centering
\caption{
Comparisons of MAL with SCL in accuracy (\%) under three dataset settings. 
}
\label{tab:negative}
\begin{tabular}{l|l|l|l} 
\toprule
\multirow{2}{*}{Loss setting} & \multicolumn{3}{l}{Dataset setting}  \\ 
\cline{2-4}
                                  & 6-60 & 10-60 & 10-100                \\ 
\hline
SCL             & 94.6 & 96.5  & 96.4                  \\
MAL (ours)              & 95.1 & 97.2  & 96.7                  \\
\bottomrule
\end{tabular}
\end{table}

\subsubsection{Visualization of Memory Addressing Vector}
\begin{figure}[htbp]
\centering
\includegraphics[width=1.0\linewidth]{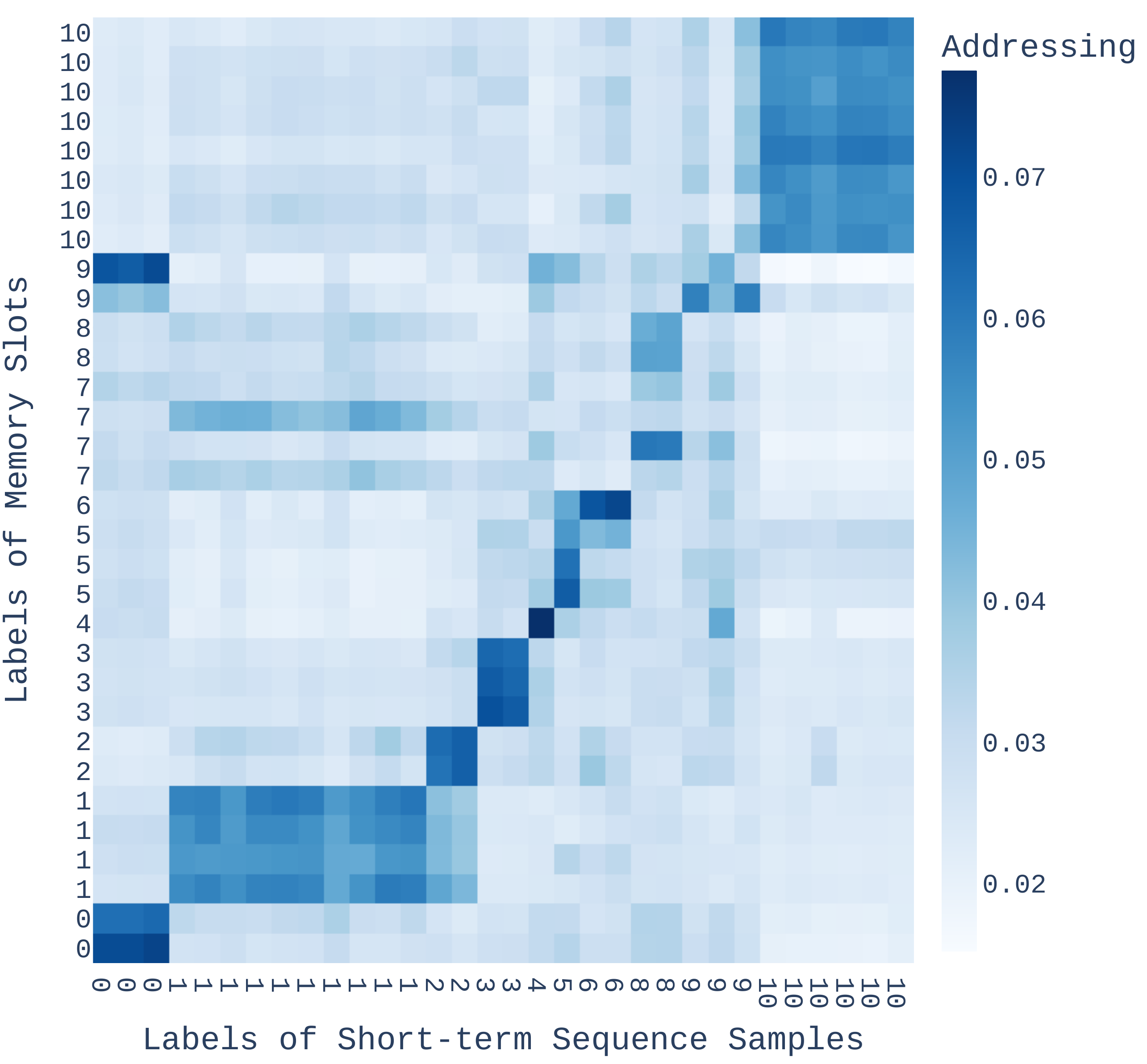}
\caption{
Visualization of memory addressing vector elements between the input short-term sequence samples and the memory slots. 
The y-axis and x-axis indicate the label indexes of the memory slots and the label indexes of short-term sequence samples, respectively. 
The color bar indicates the value of the memory addressing vector elements.
}
\label{fig:visuM}
\end{figure}

Figure \ref{fig:visuM} visualizes the memory addressing vector which represents the similarity relationship between input short-term sequence samples and memory slots in \(M\). 
For visualization clarity, we randomly select 32 memory slots and 32 short-term sequence samples and sort them by label. 
The y-axis and x-axis in Figure \ref{fig:visuM} indicate the label indexes of the memory slots and the label indexes of the input short-term sequence samples, respectively. 
The inner elements represent memory addressing vector elements between the input short-term sequence samples and the memory slots, and the more blue the color is, the more similar they are.

From Figure \ref{fig:visuM}, the features of input short-term sequence samples are highly similar to the memory slots of the same class (demonstrated as addressing vector elements with the same label index on the x-axis and y-axis have a more blue color) and have low similarity to the memory slots of different classes, which suggests that the long-term gestures sequence features stored in our memory queue \(M\) can be recalled by short-term sequence samples through similarity matching. 

\subsubsection{Confusion Matrix and Failure Cases}
\begin{figure}[h]
\centering
\includegraphics[width=0.96\linewidth]{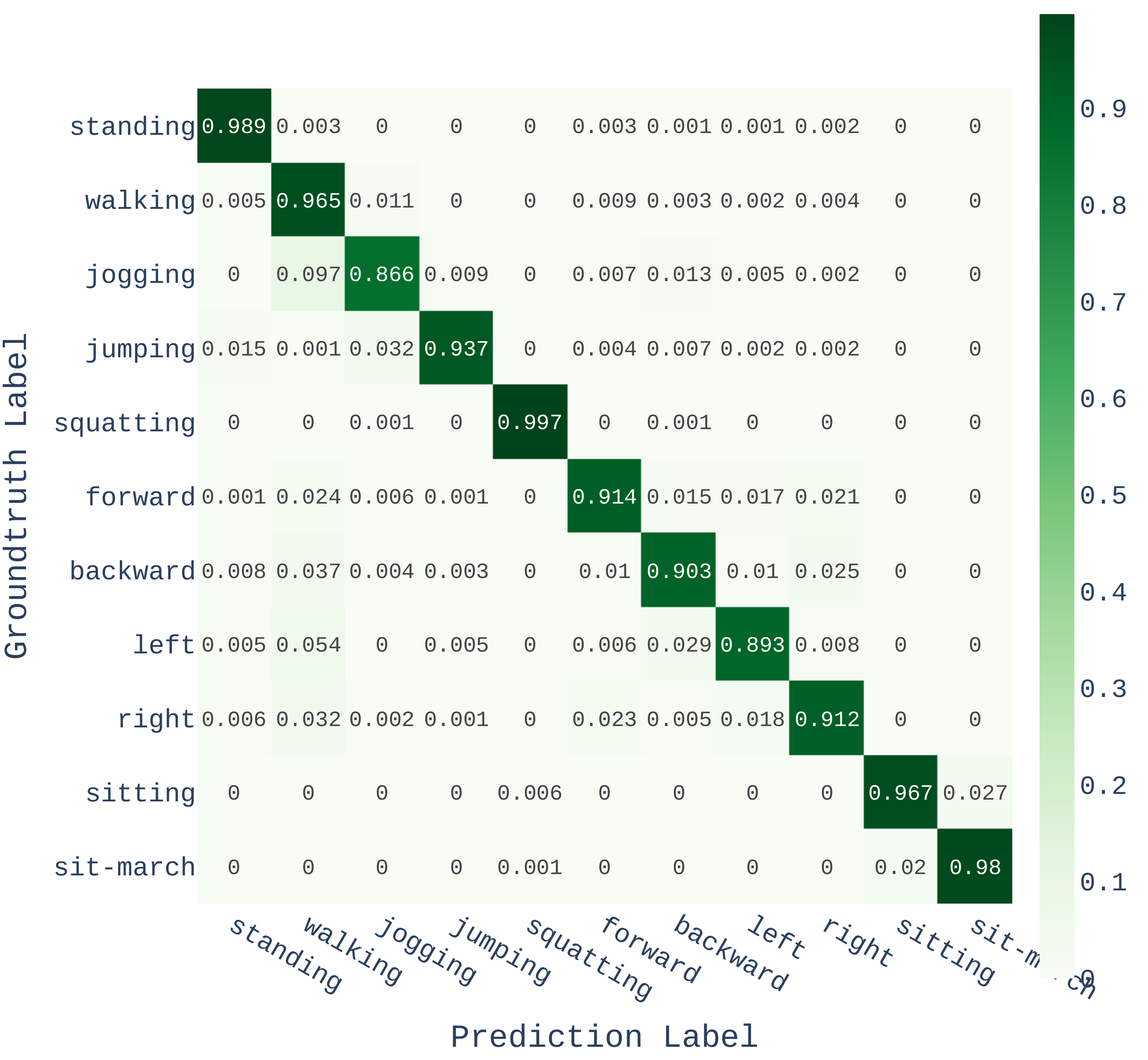}
\caption{
Confusion matrix normalized over the true condition (rows) of the model under the 6-60 dataset setting.
}
\label{fig:cm}
\end{figure}

Figure \ref{fig:cm} shows the confusion matrix of our model under the 6-60 dataset setting. 
The confusion matrix is normalized over the true condition, hence the diagonal elements of the matrix represent the recall values. 
The overall accuracy of this model is 95.1\% as shown in Table \ref{tab:ablation}. 
We notice that gestures with a recall below 91\% include jogging (86.6\%), stepping left (89.3\%), and stepping backward (90.3\%). 
Jogging is prone to be misclassified by the model as walking,  probably due to fatigue of users during jogging which results in a lower amplitude of leg movement similar to walking. 
Stepping left and stepping backward also tend to be misclassified as walking forward. 
\zlz{
These gestures do cause a position offset compared to in-place gestures such as walking in place and jumping, etc. The model may not have good generalization to counteract this position offset, resulting in a relatively low recall for these gestures. }

\subsection{Ablation Studies}
As shown in Table \ref{tab:ablation}, to verify the effectiveness of LMAN and our memory augmented loss for gesture classification, we perform an ablation study by gradually adding these two components under two dataset settings (6-60 and 10-100). 
For the baseline model containing only EfficientGCN, we input only short-term sequence samples to the encoder without long-term sequence samples, and simply feed the latent representations to the decoder. 
For the EfficientGCN+LMAN setting in the second row, we simply set the MAL to 0. 
For the EfficientGCN+MAL setting, \(M\) is only used to calculate the MAL, without providing recalled long-term sequence features for short-term sequence samples.

The experimental results show that both LMAN and MAL are beneficial for gesture classification under these two dataset settings.
When LMAN is enabled, \(M\) provides extra long-term sequence features, thus improving accuracy intuitively. 
In addition, enabling MAL further improves accuracy compared to LMAN, which suggests that MAL facilitates the encoder to learn features from the input short-term sequence samples consistent with the long-term sequence features although relevant features are not recalled. 
The model achieves the highest classification accuracy when both LMAN and MAL are enabled. 

It is worth noting that LMAN and MAL can provide more accuracy gain under the 6-60 dataset setting compared to 10-100, which may be because the sequence length of 6 frames contains less spatio-temporal information and thus relies more on the long-term sequence features provided by \(M\).

\begin{table}
\centering
\caption{Comparison with different types of model settings under two dataset settings in terms of classification accuracy (\%). \(\Delta\) shows the gain in accuracy of different model settings compared to the baseline. }
\label{tab:ablation}
\begin{tabular}{l|l|l|l} 
\toprule
\multirow{2}{*}{Model Setting} & \multicolumn{2}{l|}{Dataset setting} & \multirow{2}{*}{\(\Delta\)}  \\ 
\cline{2-3}
                               & 10-100 & 6-60                        &                         \\ 
\hline
EfficientGCN (Baseline)                  & 94.6   & 91.5                        &             0.0/0.0            \\
EfficientGCN+LMAN              & 96.4  & 93.6                        &          +1.8/+2.1              \\
EfficientGCN+MAL               & 96.6   & 94.2                        &            +2.0/+2.7              \\
EfficientGCN+LMAN+MAL          & 96.7   & 95.1                        &             +2.1/+3.6            \\
\bottomrule
\end{tabular}
\end{table}

\subsection{Comparisons with Gesture Classification Methods}
\label{section:comparisionwithexisting}
We compare our method with different models for gesture classification, including skeleton-based model ST-GCN \cite{stgcn} and EfficientGCN \cite{song2021efficientgcn} which is used as our encoder,
and point cloud-based model PCT \cite{guo2021pct} and its variant PCT+MCD \cite{zhao2021classifying}.
We also apply LMAN and MAL to ST-GCN to validate the generalization of our method.
For PCT and PCT+MCD, we perform preprocessing to reshape the data from skeleton format to point cloud format.
All these models are trained under the 6-60 dataset setting.
Comparison results are summarized in Table \ref{tab:othermehods}.

The results show that benefiting from both the strong ability of EfficientGCN to extract skeleton features and our LMAN with MAL, our method achieves the highest accuracy of 95.1\%, which significantly outperforms the other models. 
ST-GCN achieves an accuracy of 88.1\% on our dataset.
Our LMAN with MAL only stores the latent representations from the encoder output, and is thus independent of the specific encoder.
To verify the generalization of our method to other encoders, we also apply LMAN with MAL to ST-GCN and achieve an accuracy improvement of 0.8\%.

Zhao \textit{et al.} \cite{zhao2021classifying} treated gesture sequences as point clouds and trained individual models using PCT, achieving promising results.
However, our test set contains data from multiple individuals, which requires a high generalization capability of the trained model, and PCT only yields an accuracy of 85.8\% on our dataset.
PCT+MCD bridges the domain gap between the training set and test set with unsupervised domain adaptation techniques, which accesses the unlabeled sample data of the test set and improves the accuracy by 3.5\% compared to PCT on our dataset.

\begin{table}
\centering
\caption{Comparison of different models in accuracy (\%) under the 6-60 dataset setting. }
\label{tab:othermehods}
\begin{tabular}{ll} 
\toprule
Model Setting         & Accuracy  \\ 
\hline
ST-GCN \cite{stgcn}             & 88.1     \\
ST-GCN+LMAN+MAL       & 88.9      \\
PCT \cite{guo2021pct}                  & 85.8      \\
PCT+MCD \cite{zhao2021classifying}              & 89.3     \\
EfficientGCN \cite{song2021efficientgcn}          & 91.5      \\ 
\hline
EfficientGCN+LMAN+MAL (ours) & 95.1      \\
\bottomrule
\end{tabular}
\vspace{-0.3cm}
\end{table}

\section{User Study}

\subsection{VR Scenario}
\begin{figure}[htbp]
\vspace{-0.15cm}
\centering
\includegraphics[width=0.9\linewidth]{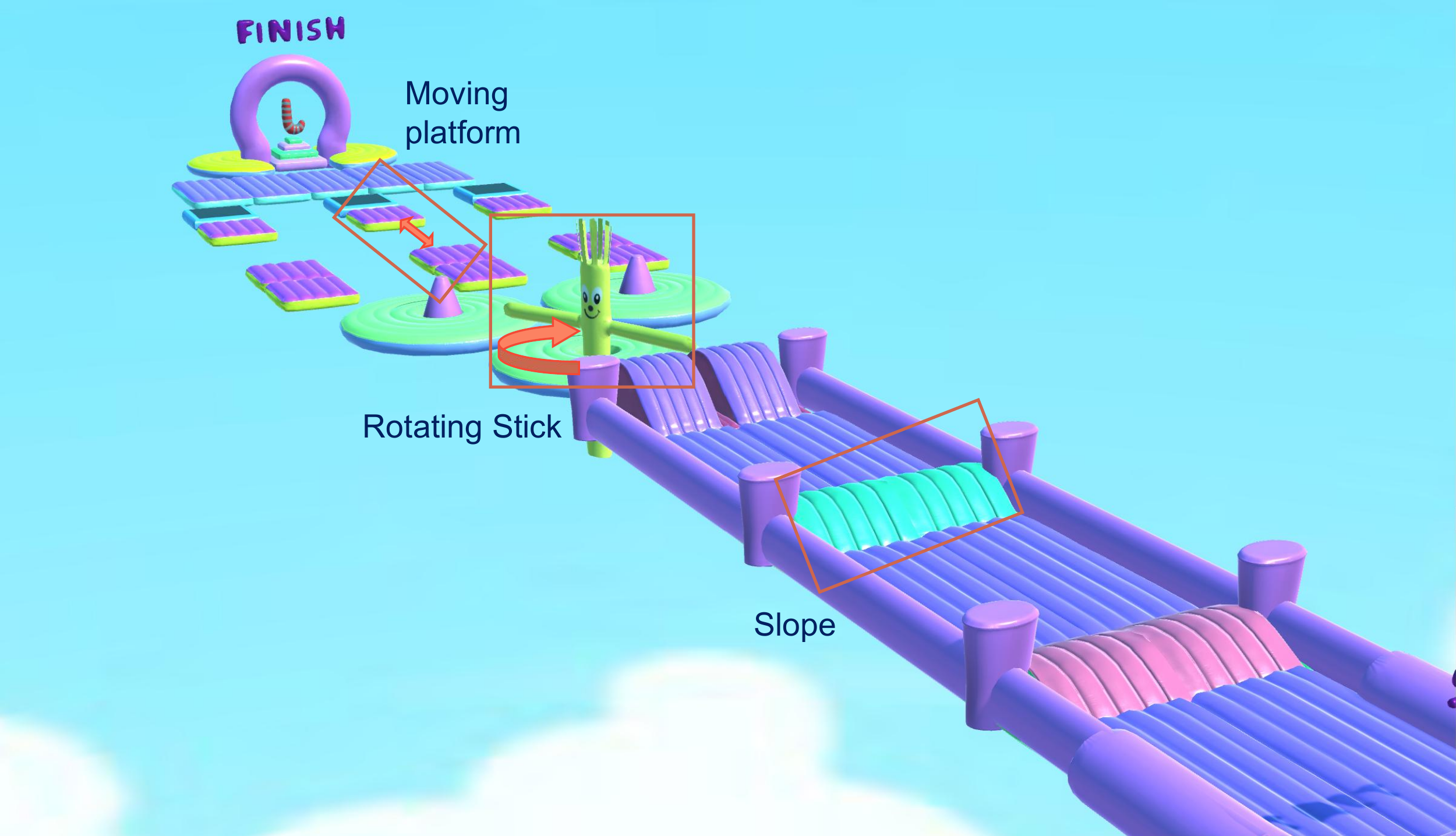}
\caption{
The virtual scenario used in our user study. 
}
\label{fig:uss}
\vspace{-0.5cm}
\end{figure}

As shown in Figure \ref{fig:uss}, we develop a parkour scenario in Unity3D\footnote{Code based on https://assetstore.unity.com/packages/templates/packs/royal-game-template-193109}, consisting of several platforms and obstacles, such as slopes, a rotating stick, and horizontally moving platforms. There are gaps between these platforms. 
Users control their virtual viewpoint from the starting point to cross different obstacles in turn.
When approaching the rotating sticks, users need to crouch or jump to avoid collisions. 
When approaching the gaps between platforms, users need to adjust their position to the edge of the platform and jump to the next platform to avoid falling.
A user departs from the starting point (i.e., bottom right corner of Figure \ref{fig:uss}), then traverses through 3 slopes, then jumps over the gaps between platforms while dodging the rotating stick, and finally arrives at the finish point.

\subsection{Experimental Setup}
\zlz{
Similar to Bowman \textit{et al.} \cite{bowman1998methodology}, our locomotion system can be summarized as: the gestures recognition module to detect gesture input and the in-game avatar controller to trigger the avatar's movement. 
}

\zlz{For the gestures recognition module,}
we deploy our trained LMAN model with the 6-60 dataset setting using Flask and expose a REST API for model inference. 
The Unity client samples HMD and trackers position data at 30Hz and predicts the user's gesture class every 180ms via HTTP requests.
\zlz{For the in-game avatar controller,}
we use the Unity Character Controller tool to control the movement of users' virtual viewpoint. 
Following Zhao \textit{et al.} \cite{zhao2021classifying}, we define the forward direction as the average value of the z-axis rotation of the two VIVE trackers.
For gestures of walking in place, stepping forward and sitting-marching, we move the viewpoint forward at a moderate speed \zlz{of 10}, and for stepping left, stepping right and stepping backward, we move the viewpoint in their corresponding direction.
Similar to walking in place, for jogging in place, we simply increase forward speed \zlz{by 1.5 times}.
As for standing or sitting, we set the moving speed to 0.
When the user is detected to be jumping, we apply a forward and upward velocity to the user's viewpoint, \zlz{which will then be cast in a parabolic path under the effect of gravity. The upward component of the jump speed is 15 and the forward component is 10.}
Finally, when the user is squatting, we scale down the user's body collider \zlz{to avoid potential head-on collisions}.

We compare our method with the PCT+MCD model trained with the same 6-60 dataset setting as our LMAN model (as described in Section \ref{section:comparisionwithexisting}), and the improved LLCM-WIP \cite{feasel2008llcm}  
\cite{cannavo2020testbed} which only recognizes two types of gestures (standing and walking in place) and therefore requires the controller to assist the input of jumping and squatting gestures with two buttons. 
We conduct the comparative user study with a within-subjects design involving these three locomotion techniques. 
The order of the three techniques is randomly assigned.

We adopt the number of times the user-controlled viewpoint falls from the platforms (denoted as Fall) and the number of times it is hit by the rotating stick (denoted as Hit) as objective metrics to measure the accuracy of virtual locomotion techniques.
The more times the Fall and Hit, the more difficult it is for the users to control the virtual viewpoint, indicating the less ease of use and effectiveness of the locomotion technique. 
Following \cite{cannavo2020testbed}, we also evaluate users' subjective perceptions of Input Responsiveness, Ease of Use, Perceived Errors and Presence with the questionnaire form  \cite{cannavo2020testbed}, which contains several questions with each question rating from 1 (Strongly Disagree) to 5 (Strongly Agree), shown in Table \ref{tab:questionnarie}. 
Similar to \cite{paik2021backward}, we interview users about their subjective perceptions of motion sickness at the end of the experiment.




\begin{table*}
\vspace{-0.3cm}
\centering
\caption{Questions to measure users' subjective feelings. \zlz{These questions are scored on a Likert scale from 1 to 5.}}
\label{tab:questionnarie}
\begin{tabular}{l|l} 
\toprule
\textbf{Subjective Metric}                   & \textbf{Question}                                                                        \\ 
\hline
\multirow{2}{*}{Responsiveness}   & The response to user input was acceptable.                                               \\ 
\cline{2-2}
                                  & The response time did not affect my
performance.                                         \\ 
\hline
\multirow{2}{*}{Ease of Use}      & I found it easy to move or reposition myself in the
virtual environment.                 \\ 
\cline{2-2}
                                  & I found it easy to undo mistakes and return to a
previous state.                         \\ 
\hline
\multirow{2}{*}{Perceived Errors} & The interfaces provided protection against
trivial errors.                               \\ 
\cline{2-2}
                                  & The interface was very robust and reliable.                                              \\ 
\hline
\multirow{2}{*}{Presence}         & I got a sense of presence, i.e., of ``being there'' during the experience.                  \\ 
\cline{2-2}
                                  & I had a good sense of scales while moving and interacting with the virtual environment.  \\
\bottomrule
\end{tabular}
\vspace{-0.2cm}
\end{table*}

\subsection{Study Procedure}
We recruited 12 participants (3 females, 9 males) from a local university with an average age of 23.5 (SD: 2.1).
Four participants reported that they had only heard of the concept of VR and never tried VR applications. Six participants said they had used VR occasionally. Two participants were familiar with VR. 
As with Paik \textit{et al.} \cite{paik2021backward}, the study procedure is as follows. 

\begin{itemize}
\item We investigate participants' familiarity with VR using the questionnaire from \cite{shi2019accurate}. 
\item Participants watch the video of the study procedure, then wear the HMD and trackers and perform a short tutorial to experience the 11 gestures. 
\item Participants complete the parkour scenario task using three locomotion techniques in a random order, with each technique running for approximately 3 minutes. 
\item After completing the scenario task for each technique, they answer the questionnaire in Table \ref{tab:questionnarie} and describe their subjective perceptions on motion sickness.
\end{itemize}

\subsection{Results}

\begin{figure}[htbp]
\centering\includegraphics[width=0.95\linewidth]{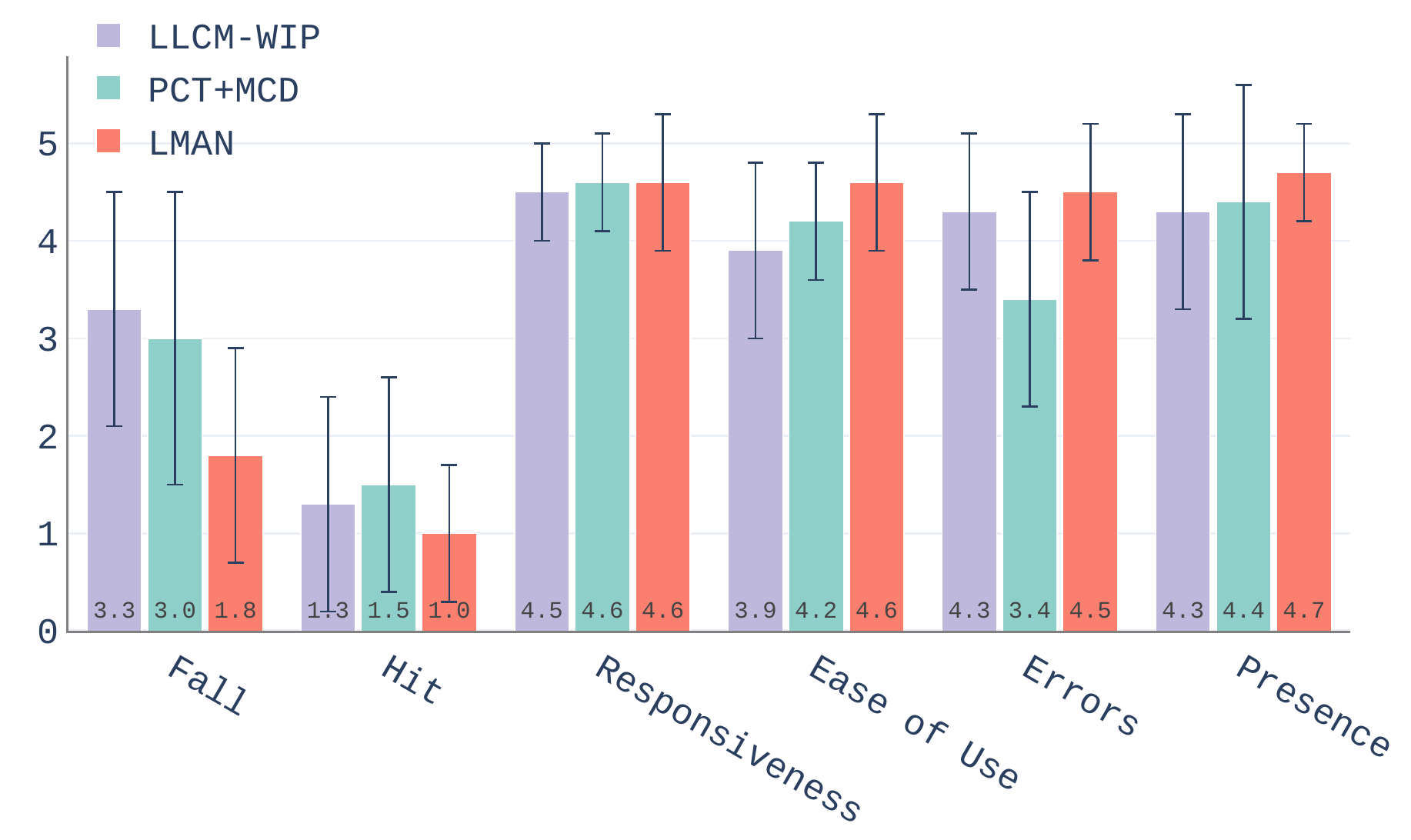}
\caption{
User study results of Fall, Hit, Responsiveness, Ease of Use, Perceived Errors and Presence metrics for the three locomotion techniques. 
}
\label{fig:usresult}
\vspace{-0.3cm}
\end{figure}

The results of the Fall, Hit, Responsiveness, Ease of Use, Perceived Errors and Presence metrics for the three locomotion techniques are presented in Figure \ref{fig:usresult}. Since there are multiple questions for each subjective metric, we use the average score of each question as the result of the metric.
\zlz{Following previous works \cite{paik2021backward,ke2021larger},} 
we perform the analysis of variance (ANOVA) and Tukey posthoc tests for technique comparisons.

The ANOVA results show that significant differences exist in Fall (\(p<0.01\)) and Perceived Errors (\(p<0.01\)). 
We can observe that the Fall metric of the LLCM-WIP technique is the highest, with an average value of 3.3 \zlz{(SD = 1.2)}, which is significantly greater than that of LMAN \zlz{(Mean = 1.8, SD = 1.1) by 1.5 (\(p<0.05\)).}
\zlz{The Fall metric of PCT+MCD technique \zlz{(Mean = 3.0, SD = 1.5)} is also greater than that of our LMAN by 1.2, showing marginally significant difference (\(p=0.08\)).}
This may be because LLCM-WIP can only recognize two gestures (walking in place and standing). So the users can only control the viewpoint to move forward, but can not move left, right or backward. 
We notice that if the users want to adjust their position when they are close to the edge of the platforms, they have to turn their body to adjust the direction first, which tends to cause the system to misidentify the body turn as walking in place, thus resulting in the users falling off the edge. 
While our LMAN can accurately identify footsteps in all four directions, thus avoiding falling. 
\zlz{For Perceived Errors, we find LMAN achieves the highest average score of 4.5 \zlz{(SD = 0.7)}, which is significantly greater than that of PCT+MCD \zlz{(Mean = 3.4, SD = 1.1)} by 1.1 (\(p<0.05\)) and significantly greater than that of LLCM-WIP \zlz{(Mean = 4.5, SD = 0.7)} by 0.9 (\(p<0.05\)).}
The PCD+MCD technique has an accuracy of only 89.3\% under the 6-60 dataset setting, and the frequent misidentified gestures significantly degrade the user experience, resulting in lower Perceived Errors scores. 
\zlz{For Ease of Use, LMAN \zlz{(Mean = 4.6, SD = 0.7)} is greater than LLCM-WIP \zlz{(Mean = 3.9, SD = 0.9)} by 0.7, showing a marginally significant difference (\(p=0.08\)).}
\zlz{For Responsiveness, LMAN (Mean = 4.6, SD = 0.7), PCT+MCD (Mean = 4.6, SD = 0.5) and LLCM-WIP (Mean = 4.5, SD = 0.5) show no significant differences (\(p>0.05\)),} 
which suggests that compared to the LLCM-WIP technique that has almost no latency, the latency of our LMAN is acceptable and does not negatively affect the user experience.

At the end of the experiment, we interviewed users of their feelings of motion sickness and encouraged them to speculate on gestures that might contribute to motion sickness. 
For the LLCM-WIP condition, 6 participants (50\%) mentioned feeling dizzy when squatting and jumping with the controller button (e.g., \textit{``I feel squatting and jumping while I am standing, which makes me feel strange and a little dizzy.''}).
This may be due to the inconsistent visual and physical perception of the participants during controller-driven jumping and squatting.
For the PCT+MCD condition, 5 participants (42\%) reported that they felt dizzy when the system incorrectly recognized gestures, for example, the system incorrectly detected a backward step and moved the virtual viewpoint backwards while the participant remained walking in place. 
This requires a high accuracy of the gestures classification model to avoid frequent mis-recognition of gestures.
For the LMAN condition, 4 participant (33\%) reported motion sickness during jogging in place (e.g., \textit{``I felt that the virtual viewpoint was moving faster than my stepping speed, and this speed mismatch made me feel a little dizzy.''}).

In summary, benefiting from our LMAN's high classification accuracy and low latency, it achieves significantly high scores in Fall and Perceived Errors metrics, and comparable scores in Hit, Ease of Use and Presence metrics. 
In addition, LMAN requires the setting of appropriate parameters to match the virtual and physical locomotion speed to avoid motion sickness.

\section{Discussion}
In this paper, we propose a novel long-term memory augmented network for in-place gestures  classification using sensor sequence data from HMD and two trackers. 
We verify the effectiveness and responsiveness of the model with a parkour game scenario. 
In this section, we first discuss the difference between the skeleton format and the point cloud format of the sensor data,
and then discuss limitations and future research on in-place gestures classification. 

\subsection{Comparison of Sensor Data Formats: Skeleton or Point Cloud}
Motion sequence data collected by motion capture sensors such as HMD and trackers is available in a number of formats for deep learning networks. 
Paik \textit{et al.} \cite{paik2021backward} designed the sensor data as the two-dimensional temporal sequence (i.e., \(x_{2d} \in \mathbb{R}^{T \times CV}\)). 
Zhao \textit{et al.} \cite{zhao2021classifying} treated several consecutive frames as a point cloud (i.e., \(x_{pc} \in \mathbb{R}^{C \times TV}\)), with HMD and two VIVE trackers providing three points in each frame. 
In this work, we treat sensor sequence data as skeleton sequences (i.e., \(x_{sk} \in \mathbb{R}^{C \times T \times V}\)), with each frame of skeleton consisting of three joints (i.e., head and left and right thighs) from the HMD and two VIVE trackers, where the left and right thigh joints are connected to the head joint with edges. 
Here we focus on the differences between the skeleton format and the point cloud format. 

Compared to point clouds, a distinctive characteristic of the skeleton is that the skeleton can be represented as a spatio-temporal undirected graph, with the joints of the skeleton represented by graph nodes \cite{stgcn}. 
The intra-body edges between body joints in each frame contain edge features including angles and lengths, and the inter-frame edges connect the same joints between consecutive frames contain joint features of inter-frame differences, e.g., the motion speed of the joint. 
Manually designing features such as motion velocity and edge length, edge angle, etc.,  before inputting raw skeleton coordinate features to the network can improve the ability of the network to extract skeleton features, which shows the importance of these skeleton features  \cite{song2021efficientgcn}. 
On the other hand, a point cloud is a set of unordered 3D points, which is the most straightforward and simple way to represent 3D geometric information \cite{guo2020deep}. 
The point cloud mixes the points of all frames of the motion sequence data in the geometric space, thus hiding features of the temporal dimension, such as the velocity or acceleration. 
In addition, the features of intra-body edges between joints that represent the logical relationship of human body are also hidden. 
The results in Table \ref{tab:othermehods} show that the accuracy of the ST-GCN \cite{stgcn} model using raw skeleton data exceeds that of the PCT \cite{guo2021pct} model using raw point cloud data by 2.3\%. 
Therefore, the skeleton format may be more suitable than the point cloud format for motion sequence data and gestures classification. 
However, this conclusion needs to be further verified by conducting more extensive experiments in the future. 

In summary, 
there are usually only a few wearable sensors available  when capturing motion data for VR applications \cite{hanson2019improving,zhao2021classifying,paik2021backward}, resulting in the samples containing fewer points for the point cloud or joints for the skeleton. This suggests that the samples contain less spatial information, and the skeleton format with extra edge features is a good choice for gestures recognition.

\subsection{Limitations and Future Work}
While our study presents an effective deep learning framework to enable in-place gestures classification with low latency, we also identify limitations for future work. 
First, we collect 25 participants' motion data (over 1.5 million frames), 
and manually labeled each frame, which is quite time-consuming and tedious, and refrains us from building a larger dataset. 
\zlz{In addition, our current method only allows the users to perform some predefined gestures within the dataset.}
In the future, we plan to leverage unsupervised learning or few-shot learning 
\zlz{to allow the recognition of new user-defined gestures}. 
Second, we focus on accurately and quickly classifying in-place gestures for virtual locomotion, while ignoring factors such as walking or jogging speed, jump height, etc. These are also important for controlling virtual viewpoint movement, and in future we would like to incorporate them into our framework. 

\section{Conclusion}
In this paper, we propose a novel Long-term Memory Augmented Network for classifying 11 in-place gestures for virtual locomotion. 
Our LMAN involves an external memory queue to store long-term sequence features, which can be recalled by short-term sequence features to provide extra contextual information. 
In addition, we design the memory augmented loss to encourage LMAN to memorize more relevant and robust features. 
Experimental results show that our method achieves a promising accuracy of 95.1\% with a latency of 192ms, and an accuracy of 97.3\% with a latency of 312ms. User study also confirms the effectiveness of our approach. 

%


\bibliographystyle{abbrv-doi}

\bibliography{template}
\end{document}